\documentclass[prb,aps,twocolumn,showpacs]{revtex4}

\usepackage{amsmath}
\usepackage{graphicx}
\usepackage{color}
\usepackage[centerlast,small]{caption}
\usepackage{subfig}
\DeclareMathAlphabet{\mathpzc}{OT1}{pzc}{m}{it}

\begin{document}

\title{Giant reflection band and anomalous negative transmission in a resonant dielectric grating slab: application to a planar cavity}
\author{L. Pilozzi, D. Schiumarini, N. Tomassini, A. D'Andrea}
\address{Istituto dei Sistemi Complessi, CNR, C.P. 10, 
Monterotondo Stazione, Roma I-00015}
\date{\today }

\begin{abstract}
  The fundamental optical effects that are at basis of giant reflection band and anomalous negative transmission in a self-sustained rectangular dielectric grating slab in P polarization and for incidence angle not very far from the Brewster's angle of the equivalent slab, are investigated.
Notice, that the self sustained dielectric grating slab is the simplest system that, due to the Bragg diffraction, can show both the former optical effects. 
A systematic study of its optical response is performed by an analytical exact solution of the Maxwell equations for a general incidence geometry.

At variance of the well known broad reflection bands in high contrast dielectric grating slab in the sub-wavelength regime, obtained by the destructive interference between the travelling fundamental wave and the first diffracted wave (a generalization of the so called second kind Wood's anomalies), the giant reflection band is a subtle effect due to the interplay, as well as among the travelling fundamental wave and the first quasi-guided diffracted one, also among the higher in-plane wave-vector components of the evanescent/divergent waves.

To better describe this effect we will compare the optical response of the self-sustained high contrast dielectric grating slab with a system composed by an equivalent homogeneous slab with a thin rectangular high contrast dielectric grating engraved in one of the two surfaces, usually taken as a prototype for the second kind Wood's anomalies generation.

Finally, the electromagnetic field confinement in a patterned planar cavity, where the mirrors are two self-sustained rectangular dielectric grating slabs, is briefly discussed.

\end{abstract}

\maketitle

\section{Introduction}
 
Diffractive phenomena in dielectric materials are at basis of many interesting optical effects shown by the light propagation in complex systems. The so called Wood's anomalies\cite{wood,Rayleigh}, the super-radiant effect\cite{Dicke,Agranovich}, band gaps in photonic crystals\cite{Yablonovich}, anomalous propagation in left handed materials\cite{Luo} and light localization in amorphous photonics\cite{John} are some of the most interesting properties that take a crucial role in the tailoring of the optical devices, allowing a full control of the local photon density of states.

While the resonant anomalies of optical diffraction gratings have been observed at the very beginning of modern optics, when they were studied by Wood and Rayleigth in the Philosophical Magazine\cite{wood,Rayleigh}, all the further interesting effects have been obtained recently, due to the improvement in the nano-manipulation of the samples.

It is well known\cite{Magnusson,rosen} that in the optical response study of a self-sustained dielectric grating slab (SSGS) three different photon energy zones can be considered: i) the so called equivalent slab model (ESM) zone, from zero to the threshold energy of the first diffraction wave, where only the zero order propagates, ii) the energy zone of the second kind Wood's anomalies (SKWA) that ranges from the first diffraction threshold till its escape into the vacuum (this zone is also called subwavelength energy zone), and finally iii) the zone of the first kind Wood's anomalies (FKWA) for higher energies. 

Some years ago two of the present authors pointed out that in the SKWA energy range, a very broad reflection band (giant reflection band: GRB) can be obtained in a self-sustained rectangular dielectric grating slab in P polarization for an incidence angle close to the Brewster angle of the ESM \cite{laura1,fenn}. They showed that this interesting property appears in the presence of a  complex coupling among travelling, guided, evanescent and divergent electromagnetic waves.

Recently, high contrast sub-wavelength self-sustained dielectric grating slabs at normal incidence have been used in vertical cavity surface emitting lasers (VCLSELs), in substitution of the vacuum/cavity Bragg reflector, since they show a much broad reflection band and a better response as a function of temperature with respect to the massive $\lambda$/4 Bragg reflectors\cite{Lu,Huang}.

It is worth remembering that high reflectivity in sub-wavelength dielectric grating slabs can be obtained by a destructive interference at the outgoing surface (zero transmission) in the energy range between the first and the second energy threshold of the diffracted waves (SKWA)\cite{rosen,Tikhodeev,Felbacq}.
However these high reflection bands are usually very narrow in energy except when large Fabry-Perot oscillations, due to the zero order diffraction wave, are present at low photon energies. An example are the rather broad and sharp edged reflection bands that can be obtained for S polarized light\cite{laura1,fenn}.

On the contrary for P polarization, close to the Brewster angle of the equivalent slab, the Fabry-Perot oscillation intensities of the zero order wave are strongly depressed and giant reflectivity bands can be obtained by a synergic effect among the SKWA sharp resonances and the electric field components with high in-plane wave vectors.\cite{laura1,fenn}.

In fact, in a dielectric grating slab, at variance of the equivalent homogeneous slab, the Fourier components of the electric field with high in-plane wave vectors, can also be travelling in the direction normal to the surfaces, and a complex interplay with surface evanescent/divergent waves can be generated.

In the present work the conditions for which the giant reflection bands can be obtained will be discussed for a general incidence geometry. 
Moreover we will show that in the energy range where the first diffracted wave propagates in the vacuum, an anomalous negative transmission can be observed for a positive real dielectric tensor. For particular values of the incidence angle all the intensity can be carried out by the first transmitted order, that propagates in the vacuum with a negative wave vector\cite{du}.

The aim of the present work is twofold. First of all, we will discuss the different optical effects that are at basis of giant reflection bands formation in rectangular dielectric grating slabs for different optical polarizations. Second, by increasing the incident photon energy we will study the zone where also the first diffracted wave becomes travelling in the vacuum. In this case a negative propagation, that leads to the so called \textit{super-lenses effect}, just observed in 2D-photonic crystal slab\cite{Luo} and in 2D anisotropic waveguides\cite{Podolskiy}, can be obtained.  Finally, the property of the electromagnetic confinement between two parallel dielectric gratings in optical micro-cavities will be also briefly discussed.

The plan of the paper is the following. 

In section II we describe the theoretical framework used to model the optical response of a self-sustained dielectric grating slab.

The different optical contributions due to travelling, guided and evanescent/divergent waves, that are at basis of GRBs in dielectric grating slabs for P polarization are discussed in section III where a systematic study of the optical response for a highly symmetric dielectric tensor grating will be presented. We will point out that broad reflection bands for a general polarization of the incidence ray are a robust property of dielectric grating slabs when characteristic resonance conditions are fulfilled. Moreover, we will underline that "giant reflection bands" are obtained in P polarization close to the Brewster angle of the equivalent slab model.

In the same section we study the energy zone where the first diffracted wave becomes travelling in the vacuum and show that, for selected parameter values, negative transmission can be achieved in a slab of 1D photonic crystal\cite{du} by Bragg effects.

In section IV an interesting application of dielectric grating slabs in optical micro-cavities will be suggested and briefly discussed.

Conclusions are given in section V.

\section{Theory}
The Maxwell equations in photonic crystal slabs for S and P polarizations are usually solved as an eigenvalue problem with respect to the electric and magnetic fields.
In the present paper we choose to solve the equations as a function of the electric field alone, by magnetic field elimination, namely:
\begin{equation}
 \boldsymbol{\nabla }  \times \boldsymbol{\nabla }  \times \textbf{E} (\textbf{r}) = \frac{{\omega ^2 }}
{{c^2 }}\textbf{D}(\textbf{r})\,\,\,\,\,.
\end{equation}

We consider a general scattering geometry of an incident plane wave on the SSGS, periodic along the x-axis and with wires direction along the y-axis, as shown in Fig.1.
The plane of incidence performs an angle $\varphi _o $ with the Cartesian plane (x,z) and the angle between the wave vector, incident from the vacuum, with the z-axis is $\vartheta _o $.

We model the 1D rectangular dielectric grating\cite{fenn}, with periodicity $d$ and wires width $L_x $ along the x axis, of a non-magnetic material ($\mu _o  = 1$), with a bulk local dielectric function \[
\varepsilon (\omega ) = \varepsilon '(\omega ) + i\varepsilon ''(\omega )
\] and Fourier transformed dielectric tensor 
\begin{equation}
\varepsilon _{G,G'} (\omega ) = {\varepsilon _o \delta _{G,G'}  + \left[ {\Delta \varepsilon (\omega ) + i\varepsilon ''(\omega )} \right]f_x S(G  - G' )}\
\end{equation} 
where $f_x   = L_x  /d  $ is the filling factor, $\Delta \varepsilon (\omega ) = \varepsilon '(\omega ) - \varepsilon _o $ the dielectric contrast, $\varepsilon _o $
the vacuum dielectric constant and $ S(G  - G')$ the Fourier transformed geometrical tensor (structural factor): \[S(G  - G' ) = \frac{{\sin \left[ {(G   - G' )L_x  /2} \right]}}{{(G   - G' )L_x  /2}}\]
The reciprocal lattice vector components are: $G = \ell 2\pi /d $ and $G'  = \ell '\,2\pi /d  $ with $\ell,\ell ' = 0, \pm 1, \pm 2,..., \pm N$ and $N \to \infty $.

The limit $G\to G'$ defines the so called "equivalent slab approximation"  of the grating with dielectric function:
\begin{eqnarray}
\bar \varepsilon (\omega ) \equiv 
\lim _{G \to G'} \varepsilon _{G,G'} (\omega ) = \frac{{\varepsilon _o (d - L_x ) + \varepsilon (\omega )L_x }}
{{d }} =\nonumber\\
=(1 - f_x )\varepsilon _o + f_x \varepsilon (\omega)
\end{eqnarray}

In the energy range of low absorbtion ($\varepsilon ''(\omega ) \to 0$) and $\varepsilon '(\omega ) =\varepsilon '=const$, the resonance conditions among different in plane wave vectors in the grating slab requires:

i) a large dielectric contrast 
$\Delta \varepsilon =  {\varepsilon ' - \varepsilon _o } \geq 10$;

ii) a large filling factor $f_x  = L_x /d =0.5 \div 1.0$  that determines also the effective dielectric function $\bar \varepsilon $  and

iii) comparable slab thickness and wires width $L_z /L_x =1.0 \div 1.5$ that determine the photon Mie's resonances in each rectangular elementary cell.

Clearly the simultaneous occurrence of the ii) and iii) conditions strongly influences the building up of quasi-guided waves in the dielectric grating slab, and their interplay with the travelling fundamental one.

Moreover, while the i) and ii) conditions influence the strong coupling among the different in-plane wave vector electric field Fourier components, generated by the grating periodicity (see eq.(2) where the coupling scales as the product of the two quantities), the iii) condition determines the destructive interference between the two propagating waves (travelling and quasi-guided). 

Notice that the dielectric tensor of eq.(2), with row index $\ell $ and column index $\ell' $, is symmetric and invariant for translation along its principal diagonal, being $\bar \varepsilon $ its value; the matrix elements of all the other diagonals ($\ell' =\ell \pm m $ for $m  = \pm 1, \pm 2,... \pm M$) are proportional to the dielectric contrast value $\Delta \varepsilon $ divided the distance $m $ from the principal diagonal.
Moreover, for commensurate values between wire width and grating periodicity ($L_x /d  = \nu /n$ with $\nu$ and n integer numbers) the matrix elements of the diagonals with $m = \pm n$ are zero (see eq.2), as well as all the other elements in the diagonals with m a multiple of n.

A systematic study is performed by choosing, as duty cycle values of the rectangular dielectric grating, the lowest Pitagora's fractions p/(p+1) with integer p=1,2,3,4 values, and optimizing the $L_z/L_x$ ratio in order to obtain the giant reflection band.

The dielectric tensor simmetry affects the strong coupling among the incident zero diffraction (G=0) electromagnetic field component and the higher orders ($G \ne 0$) in the grating slab and it can be directly observed in the S-polarized optical response. In fact, for a negligible y-component of the photon wave vector ($q_y \to 0$) and in the limit $L_z \to \infty $ the Maxwell equations reduce to an eigenvalues problem:

\begin{equation}
\sum\limits_{G'} M_{G,G'}(\omega) \varphi _n (G') = k_n^2 \,\varphi _n (G)
\end{equation}
where the dynamical matrix has the same symmetry of the dielectric tensor of eq.(2) except for the matrix elements of the principal diagonal \[M_{G,G'}(\omega)= {\frac{{\omega ^2 }}
{{c^2 }}\varepsilon_{G,G'} - q_x^2 (G)\delta _{G,G'}
 }\mathop  = \limits_{G = G'} 
 {\frac{{\omega ^2 }}
{{c^2 }}\,\overline{\varepsilon} -  q_x^2 (G)}\] that are no more translationally invariant.

\begin{figure}[t]
\includegraphics[scale=0.5]{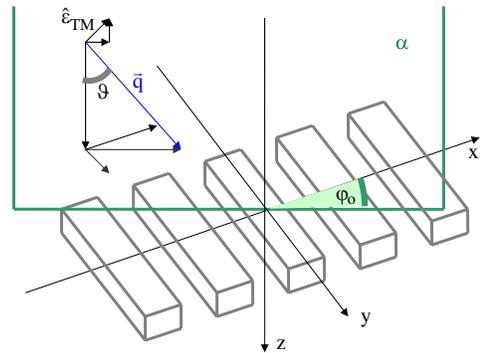}
\caption{(color online) Schematic illustration of the system under study: a self-sustained dielectric grating slab, periodic in the x direction, infinite and homogeneus in the y direction and finite in the z one.}
\label{fig:}
\end{figure}

In order to deal with finite-dimensional matrices a truncation of the diffraction orders is needed.
Considering, for example, the simple case with $\varphi_o= 0$ in S polarization, the choice of the N value to ensure the numerical convergence of the optical response calculation can be achieved by two different numerical approaches. 
We can increase the $\nu$-value ($\nu \to N$) of the square optical matrix of eq.(4) (of the order $(2\nu + 1) \times (2\nu + 1)$). In this case, we increase  both the eigenstate basis set $\left\{ {\varphi _1 (G), \varphi _2 (G) ... \varphi _{2\nu + 1} (G)} \right\}$
and the Fourier basis of the reciprocal wave vectors $\left\{ G \right\}$
 of ($2\nu + 1$)-dimension. We define this usual approach as the square matrix method (SMM).

In a second approach, that we define as the n-diagonal matrix method (nDMM), we start with a $(2N + 1) \times (2N + 1)$ dimensional square matrix in eq.(4), with only the principal diagonal matrix elements different from zero. Then, we increase symmetrically, with respect to the principal diagonal, the number of up and down diagonals with non zero matrix elements, till to reach the numerical convergence. Notice that the 1DMM defines the equivalent slab approximation (ESA).

In the first case the convergence is obtained with a Fourier truncated basis set $\left\{ {\varphi _1 (G), \varphi _2 (G) ... \varphi _{2\nu + 1} (G)} \right\}$, but with the full interaction between the different in plane components taken into account. In the second case the Fourier basis set dimension is at convergence, but with a reduced order of interactions (off-diagonal matrix elements). 
Since these two different criterions give physical complementary informations on the behaviour of photon propagation in a dielectric grating slab, we will use both in the present work.

\section{Optical response}

\subsection{Photonic modes dispersion}

Let us consider an incident electric plane wave of energy $\hbar \omega $ and wave vector $\textbf{q}$ on a surface of a rectangular SSGS that accomplish the photonic crystals condition $\varepsilon _b (\omega ) \approx \varepsilon ' _b (\omega ) \gg \varepsilon '' _b (\omega )$. Moreover, in the energy zone of very low absorption ($\varepsilon '' _b (\omega ) \to 0$) we can make the further approximation $\varepsilon ' _b (\omega ) \approx const$.

The model computation is performed  for a P polarized incident wave in the resonance conditions regime. All the necessary equations for the optical response calculation are given in Appendix A.

We use as bulk refractive index value $n_b=\sqrt {\varepsilon _b } 
 = 3.34$, filling factor $f_x  = L_x /d $
=3/4 and spatial periodicity d=300nm, while the plane of scattering is taken normal to the wires direction (the angle of scattering plane $\varphi _o =0$ in Fig.1), and the incidence angle is $\vartheta  = 60^o $ rather close to the Brewster angle of the effective dielectric function ($\vartheta  = 72^o$). The slab thickness $L_z  = 351.9nm$ ($L_z /L_x  \approx 3/2$) is optimized in order to have a strong coupling among in-plane wave vector components.

\begin{figure}[t]
\includegraphics[scale=0.55]{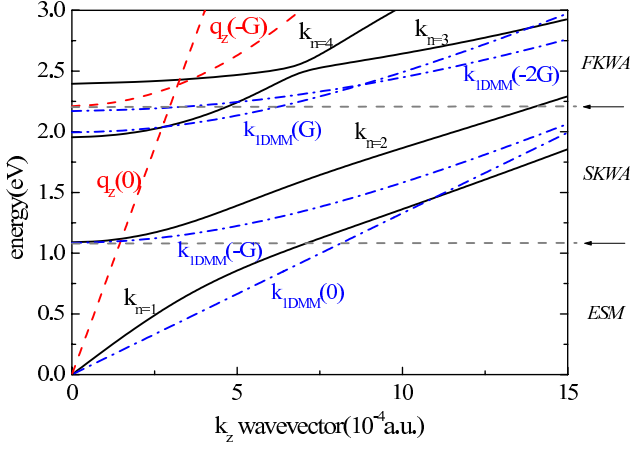}
\caption{(color online) P-polarized dispersion curves $k_n (\omega )$ (solid black line) of a grating with bulk refractive index value $n_b = 3.34$, filling factor $f_x $=3/4 and spatial periodicity d=300nm. The incidence plane is normal to the wires direction ($\varphi _o =0$), and the  incidence angle, $\vartheta  = 60^o $. The dispersion curves of the ESA (dot-dashed blu curves), as well as the two lowest dispersion curves $q_z(G)$ for G=0 and $G = - 2\pi /d_x $ in vacuum (dashed red curves) are also shown.}
\end{figure}

In the dispersion curves $k_n (\omega )$  of the dielectric grating, shown in Fig.2 (solid black line), three different energy zones, just presented in the Introduction: i) equivalent slab model (ESM), ii) second kind Wood's anomalies (SKWA), and iii) first kind Wood's anomalies (FKWA), can be easily identified in correspondence of different treshold energies.

Notice, that while the zero order curve (n=1) is travelling for any photon energy, the curves with $n > 1$ show different threshold energies (in the limit $k_z  \to 0$). For the chosen parameter values and for photon energy $\hbar \omega =1.2 eV$, a bit greater than the threshold energy of the first diffracted mode ($\hbar \omega _{1t}  \approx 1.08eV$), the corresponding wavelength, computed for the equivalent slab dielectric function $\bar \varepsilon  = f_x \,\varepsilon _b  + (1 - f_x )\,\varepsilon _o  = 8.6167$, is $\bar \lambda  = 2\pi /\bar k = 351.882nm$ rather close to the chosen slab thickness $\bar \lambda  \approx L_z $\cite{fenn}.

The dispersion curves in the ESA (dot-dashed curves), computed by taking equal zero the off-diagonal elements of the block matrices of eq.(A7), are also shown in the same picture for different diffraction orders. Notice, that the energy thresholds for the first and the second diffracted wave $G =\mp 1$ are very close to that of the grating modes (n=2,3); therefore, a rough estimation of the thresholds energies for n=2,3 diffracted waves, can be obtained by the propagation conditions of the effective homogeneous model:

\[E _ \mp   = Gc\left[ {\sqrt {\bar \varepsilon }  \mp \sqrt {\varepsilon _o } \sin \vartheta } \right]/\left[ {\bar \varepsilon  - \varepsilon _o \,\sin ^2 \vartheta } \right].\]

Moreover, since the dispersion curves $k_{1DMM}$ have been obtained by the diagonal elements of the dynamical matrix of eq.(11), the $k_{1DMM}(G)$ and $k_{1DMM}(-2G)$ curves show a crossing that is removed when the off-diagonal matrix elements are considered. 

The two lowest dispersion curves $q_z(G)$ for G=0 and $G = - 2\pi /d$ in the vacuum are also given in the same picture (dashed red curves). The G=0 line identifies, among the grating modes, the ones that are confined in the patterned slab while, for each k, the $q_z(- 2\pi /d_x) $ curve gives the treshold energy above which the -1st order can be observed in reflection and/or transmission. In the next section we will point out how the grating parameters can be choosen to enhance the -1st order transmitted intensity while depressing the zero order reflection and transmission.

Notice that the threshold energy ($\hbar \omega  = 2.21eV$) of the $G =  - 2\pi /d $ wave in the vacuum is rather close to the threshold ($\hbar \omega _{n = 3} = 1.97\,eV)$ of the second diffraction wave in the grating; moreover, while the G=0 wave emerges from the sample with the same angle of the incident one, the $G =  - 2\pi /d $ ray shows a negative propagation very sensitive to the photon energy, as will be further discussed in the next section.

\begin{figure}[b]
\includegraphics[scale=0.5]{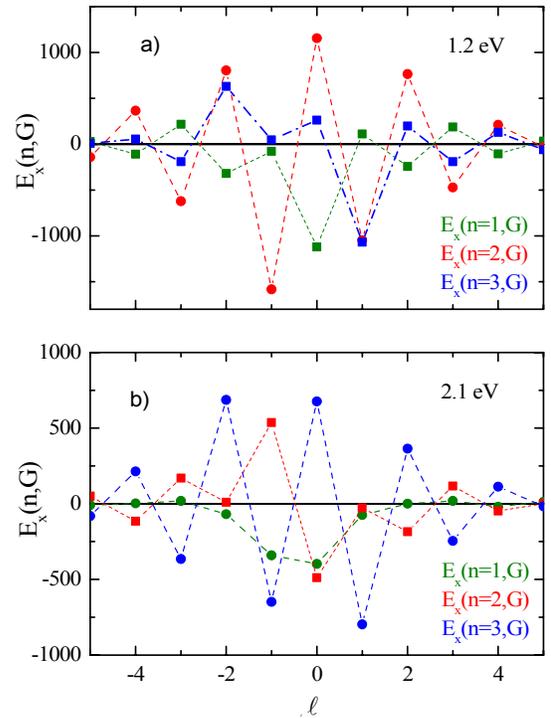}
\caption{(color online) Electric field eigenvectors $ {\varphi _n (G)}$ for n=1,2,3 computed for incidence angle $\vartheta  = 60^o $, and photon energy a) $\hbar \omega$=1.2 eV b) $\hbar \omega$=2.1 eV}
\label{fig:}
\end{figure}

In Fig. 3a) and 3b) the G-components of the electric field eigenvectors $
\left\{ {\varphi _n (G)} \right\}$ with n=1,2,3 of eq.(A7) are given for an incidence angle $\vartheta  = 60^o $, and photon energies a bit higher than the first ($\hbar \omega  = 1.2\,eV$) and the second ($\hbar \omega  = 1.97eV$) threshold energies (see Fig.2).

It is well known that for normal incidence ($\vartheta  = 0^o $) the amplitude components of the n-th order diffracted wave show even (2n+1) or odd (2n) symmetry\cite{laura1} with respect to the $G \to  - G$ transformation (obviously this property cannot be observed in the present calculation, due to the large value of the incidence angle choosen). Moreover, amplitudes distribution for n=1,2,3 shows maximum values in correspondence with $\ell  = 0, - 1, + 1$ respectively and gives an estimation of the coupling among waves with different in-plane wave vectors. The rather broad amplitude distribution around the maxima become rather sharp when the eigenvector is computed at its own threshold energy (see the distribution of n=2 and n=3 in Fig. 3a) and 3b) respectively). 

Notice that the SKWA effect in tick dielectric grating slab \cite{fenn} is a rather more complicated phenomenon with respect to that studied in planar waveguides with a thin grating engraved on one of its surfaces, a system usually considered a prototype for the SKWA explanation\cite{rosen}.

This different behaviour is schematically reported in Fig.4a) and 4b), where the propagation of the fundamental (G=0) and the first order diffraction wave ($G =  - 2\pi /d$) inside the patterned planar waveguide is compared with the propagation of the correspondent in-plane components (n=1,2) in a SSGS. Notice, that the SKWA of the two systems are qualitatively different; in fact, while on the bottom surface of the grating slab (Fig.4b) the first diffraction wave amplitude (n=2 and $G  = - 2\pi /d$) is partially transmitted by the conversion on the fundamental wave component (G=0), due to the boundary conditions, on  the planar waveguide case (Fig.4a) an internal total reflection is observed. Moreover, while the SKWA in the structure of Fig.4a) is given by the destructive interference between the fundamental ($G = 0$) and the first diffracted wave ($G =  - 2\pi /d$), in the structure of Fig.4b) many components, with the former two in plane wave vectors, interfere; therefore, we expect that the two structures show, in the optical response spectra, rather the same resonant energies (SKWA) but different line-shapes.
\begin{figure}[t]
\includegraphics[scale=0.8]{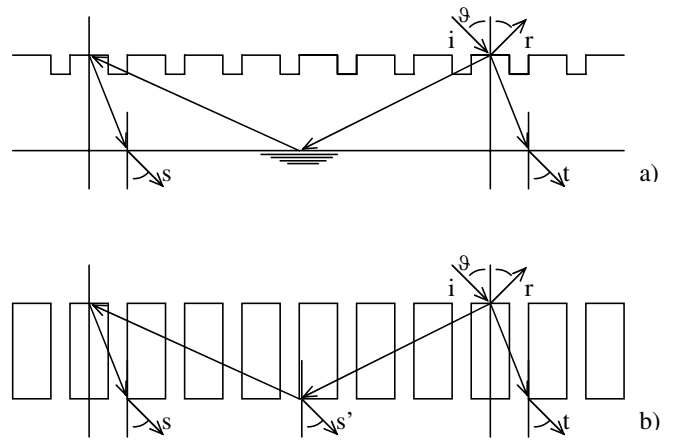}
\caption{ Schematic illustration of a) the geometry for the thin grating engraved on an homogeneous slab and b) the self-sustained grating slab (SSGS) configuration.}
\label{fig:}
\end{figure}
As an example let us consider the case of photon energy $\hbar \omega  \approx 1.242eV$ and in plane wave vector $q_x (0) = 2.8854 \cdot 10^{ - 4} a.u.$ ($\vartheta  = 60^o $) where two diffracted waves propagate in the grating with wave vectors along the z-axis $k_2  = 3.37838 \cdot 10^{ - 4} a.u.$ and $k_1  = 8.64865 \cdot 10^{ - 4} a.u.$ (as can be obtained from Fig.2). Since the two diffracted waves have their highest component amplitudes at $G_{ - 1}  =  - 2\pi /d $ and $G_0  = 0$ respectively (see Fig.3a) it is possible to determine the propagation angles inside the grating slab that in this case are $\varphi _1  \approx 18^o 30'$ and $\varphi _{ - 1}  \approx  - 67^o 30'$ respectively (see Fig. 4b) for lattice wave vector value $G_1  = 2\pi /d = 11.08354 \cdot 10^{ - 4} a.u.$ Therefore, the phase shift associated with path length difference  between incident wave and the quasi-guided wave in the grating slab gives the relation: $tg\left( {\varphi _{ - 1} } \right) = \frac{{q_x ( - G)}}  {{k_2 }} \approx m\frac{d} {{L_z }}$ with integer m. 
A more quantitative analysis of the optical response differences between the structures of Fig.4a) and 4b) will be fully discussed in the next section.
 
\begin{figure}[t]
\includegraphics[scale=0.55]{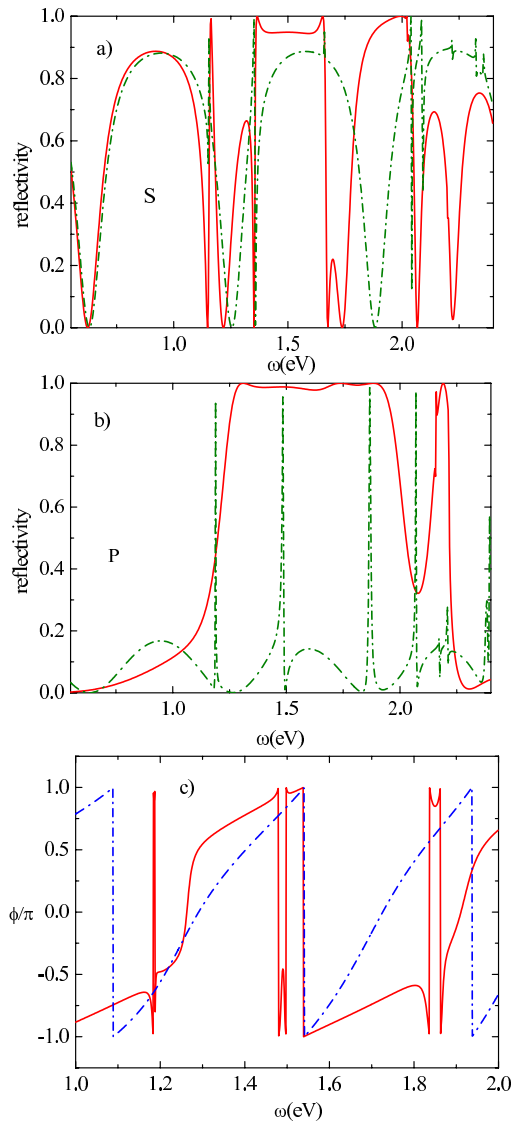}
\caption{ Reflected intensity, as a function of energy, for a self-sustained grating of thickness $L_z=351.9 nm$ (solid red curve) and for a thin grating ($L_z/10$) engraved on a thick slab ($9L_z/10$) of dielectric function $\overline{\varepsilon}$ (dot-dashed green curve): a) S polarization and  b) P polarization for incidence angles $\varphi_o=0$ and $\vartheta  = 60^o$. c) reflection phase for the P polarization case, for the engraved (solid red curve) and the self-sustained grating (dot-dashed blue curve) structure.}
\label{fig:}
\end{figure}

\begin{figure}[b]
\includegraphics[scale=0.6]{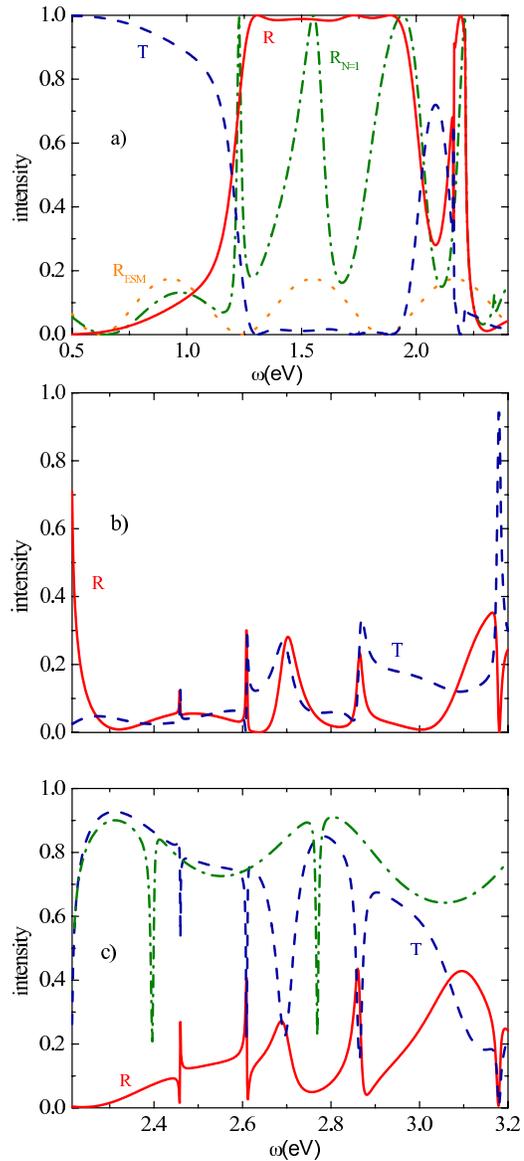}
\caption{(color online) P-polarized reflected (solid red line) and transmitted (dashed blu line) intensity for $\varphi_o  = 0^o ,\vartheta  = 60^o $, for a) the G=0 wave  in the energy range from zero to the second energy threshold and, in the FKWA range, b) the G=0 wave, c) the first diffracted one. In a) the reflection curve, $R_{ESM}$, of the ESM (dotted orange curve) and the reflection peaks $R_{N=1}$ of the SKWAs (dot-dashed green line) are also shown. In c) the transmittivity computed considering only the $G=0,\pm 2\pi /d$ components in the dielectric tensor is shown (dot-dashed green line).}
\label{fig:}
\end{figure}

\subsection{Reflectivity and transmittivity}

We would like to remind that in a dielectric grating supported on an homogeneous slab (see Fig.4a) the optical properties (reflectivity and transmittivity of the fundamental and diffractive waves) are led by two different Wood's anomalies\cite{rosen}, namely:  diffractive anomalies (or FKWA) due to the propagation of  deflected rays at vacuum sides of the system and wave guiding anomalies (or SKWA), due to the interference between incoming light and first diffracted wave, guided  into the slab (notice, that these two effects are given in a decreasing photon energy order).

\subsubsection{Giant reflection band and negative propagation}

In order to compare the optical properties of the two structures of Fig.4a) and 4b) we choose, for the engraved grating of Fig.4a), the same parameter values of the self-sustained one, with a grating thickness ($\Delta  = L_z /10$) one order of magnitude lower than the SSGS, and the homogeneous slab thickness $L = L_z  - \Delta $ with a dielectric constant $\varepsilon _s  = \bar \varepsilon $. Notice that the former parameter values are chosen in order to observe rather the same Fabry-Perot and SKWA effects in the two structures.

Moreover, while in the former structure (Fig.4a) the source of the diffracted waves and of the quasi-guided waves is given essentially to the periodic grating and the homogeneous slab respectively, in the latter structure (Fig.4b) the two effects are completely mixed in all the slab volume.

In Fig.5a) and 5b) the optical reflection spectra, for an incidence angle $\vartheta  = 60^o $ and in the SKWA energy range (see Fig.2), for both the structures of Fig.4a) and 4b) are shown for S and P polarization respectively. In Fig.5a) we observe that the optical spectra for S polarization show Fabry-Perot oscillations, for photon energies lower than the first treshold, very similar for the two different structures, and 
the SKWAs at rather the same energies, has hypothesized before\cite{fenn}. At variance, for P polarization (Fig.5b) narrow and  broad reflection bands are observed for the engraved and the self-sustained grating structure respectively. In this case, since the angle of incidence is very close to the Brewster angle of the equivalent slab model, the Fabry-Perot oscillation amplitudes are rather small for the engraved system while are completely suppressed for the SSGS and they show very different reflection spectra.

For the SSGS any possible anomalous behaviour, in the band where the reflectivity intensity remains equal to unity, can be identified only in the phase of the reflected wave. 
To this end the reflection phase versus energy, for the P polarization case, is shown in 
Fig. 5c) for both the engraved and the self-sustained grating structure. It can be seen that, as expected, the reflection phase for the engraved structure has a change equal to 2$\pi$ at the reflection peak position. The changes in the reflection phase of the self-sustained grating structure are a signature of similar resonances underlying the giant reflection band.

Now we consider the optical response for P polarization of the SSGS in the same energy range of Fig. 2. The optical spectra for reflected, transmitted and deflected rays are shown in Fig.6.   
In Fig.6a) is shown the optical response in the energy range from 0.5 eV to the energy where the first diffraction wave (G=$-2\pi /d$) escapes in to the vacuum (see dispersion curves of Fig.2). 
The three reflection curves correspond to three different basis truncations in the reciprocal lattice wave vector space, namely: i) for G=0, the reflection curve, $R_{ESM}$, of the ESM is obtained, ii) for $G = 0, \pm 2\pi /d$ the reflection peaks $R_{N=1}$ of the SKWAs are shown and finally iii) at convergence $(G = \ell \,2\pi /d$ for $\ell  = 0, \pm 1, \pm 2,..., \pm N$ for $N \approx 10)$, where also surface waves are present and the giant reflection band appears. The transmittivity is also shown in the same picture in order to check the computation 
accuracy (R+T=1).

In the case i) we observe Fabry-Perot oscillations of the equivalent slab model very similar to that of the engraved system shown in Fig. 5b), for photon energies lower than the first treshold. In the ii) case four well shaped SKWAs appear, distributed in all the energy range between first and second threshold energies, rather similar to those observed in the engraved dielectric slab (see Fig.5b). Therefore, at this degree of approximation, the similarity between the two different structures of Fig. 4a) and 4b) is recovered, with the difference that, at variance of the engraved grating structure, in the self-sustained one the SKWA peaks are very sharp in correspondence of the energy thresholds (n=2  $\hbar \omega  \sim $
1.25eV, n=3  $\hbar \omega  \sim $
1.97 eV),  while rapidly broadens for energy far from the thresholds, due to the interference of many in-plane components (see Fig.3a and 3b).  

Finally, when the numerical convergence is reached iii), a giant reflection band is obtained close to the Brewster angle value with the crucial contribution of higher order diffracted waves\cite{fenn}. For photon energy $\hbar \omega  \approx 2.2eV$ rather close to the treshold of the second diffracted wave and to the escape into the vacuum of the first diffracted wave, the synergic effect between SKWAs and higher order diffracted waves, that is at basis of the giant reflection band, deteriorates and the reflection band disappears.

The mechanism that undergo giant reflection bands in P polarization, near the Brewster angle, is then rather different from that observed for broad reflection bands in S polarization (or in P polarization for an incidence angle rather far from the Brewster angle). In the S polarization case, from the first to the second energy threshold, well shaped reflection bands originate due to the interference between the large Fabry-Perot oscillations, in the slab resonance condition, and the quasi-guided first diffracted wave (SKWA); higher order waves seem to give a minor contribution to the reflection effect (see ref. [10] and Fig.5a).

 \begin{figure}[t]
\includegraphics[scale=0.52]{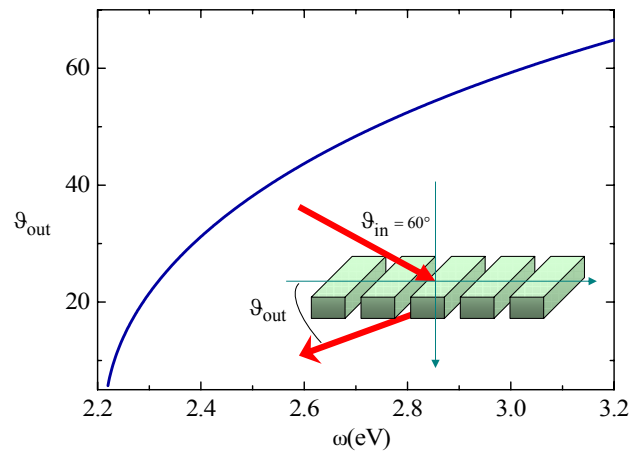}
\caption{(color online) Output angle $\vartheta _{out} $ of the -1st diffracted wave as a function of energy.}
\label{fig:}
\end{figure}

Now we move to the energy zone of the first kind Wood anomalies and show, in Fig. 6b) and 6c), the reflectivity and transmittivity intensities of the fundamental mode G=0 and the -1st diffracted one that propagates in the vacuum.

\begin{figure*}
  \centering
  {\includegraphics[scale=0.7]{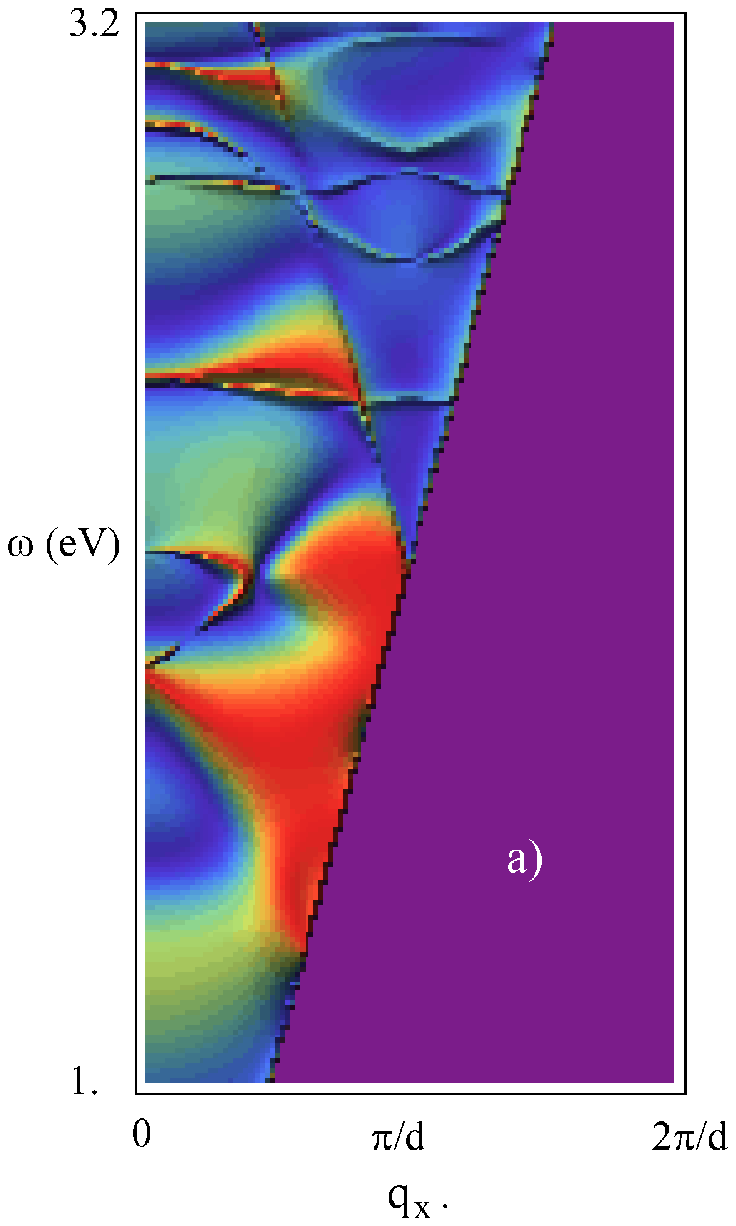}}                
  {\includegraphics[scale=0.7]{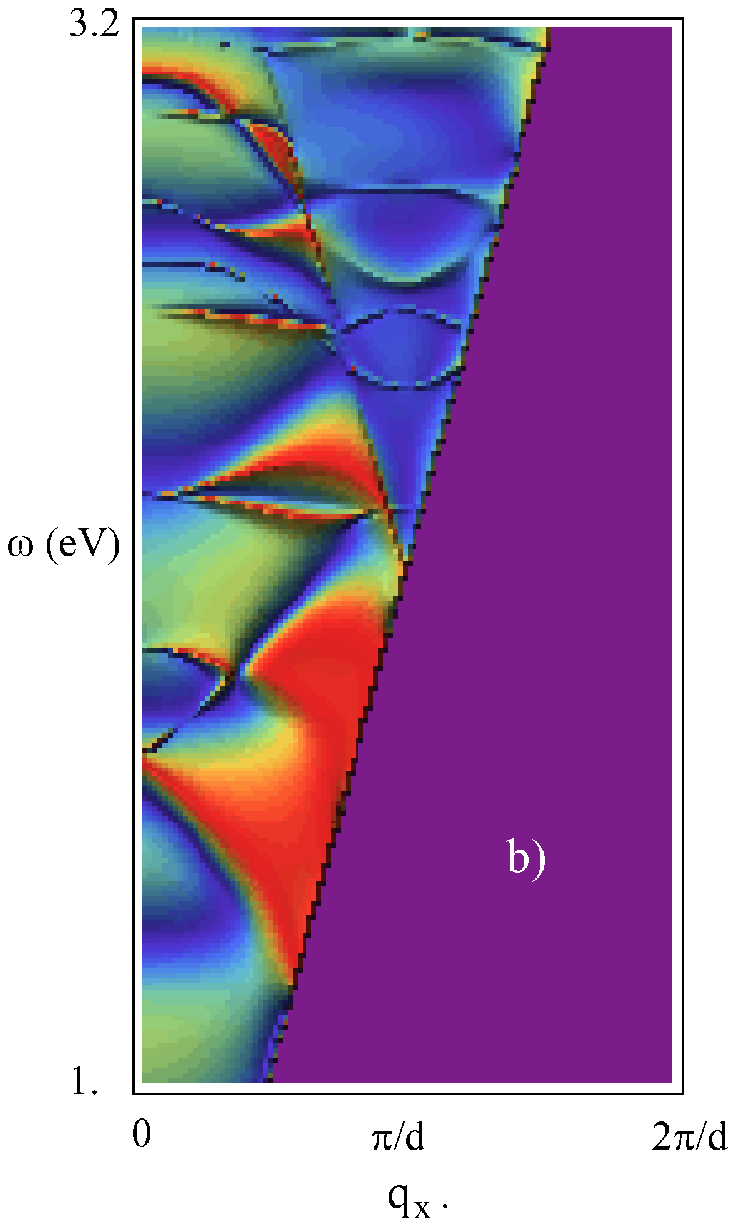}}
 {\includegraphics[scale=0.7]{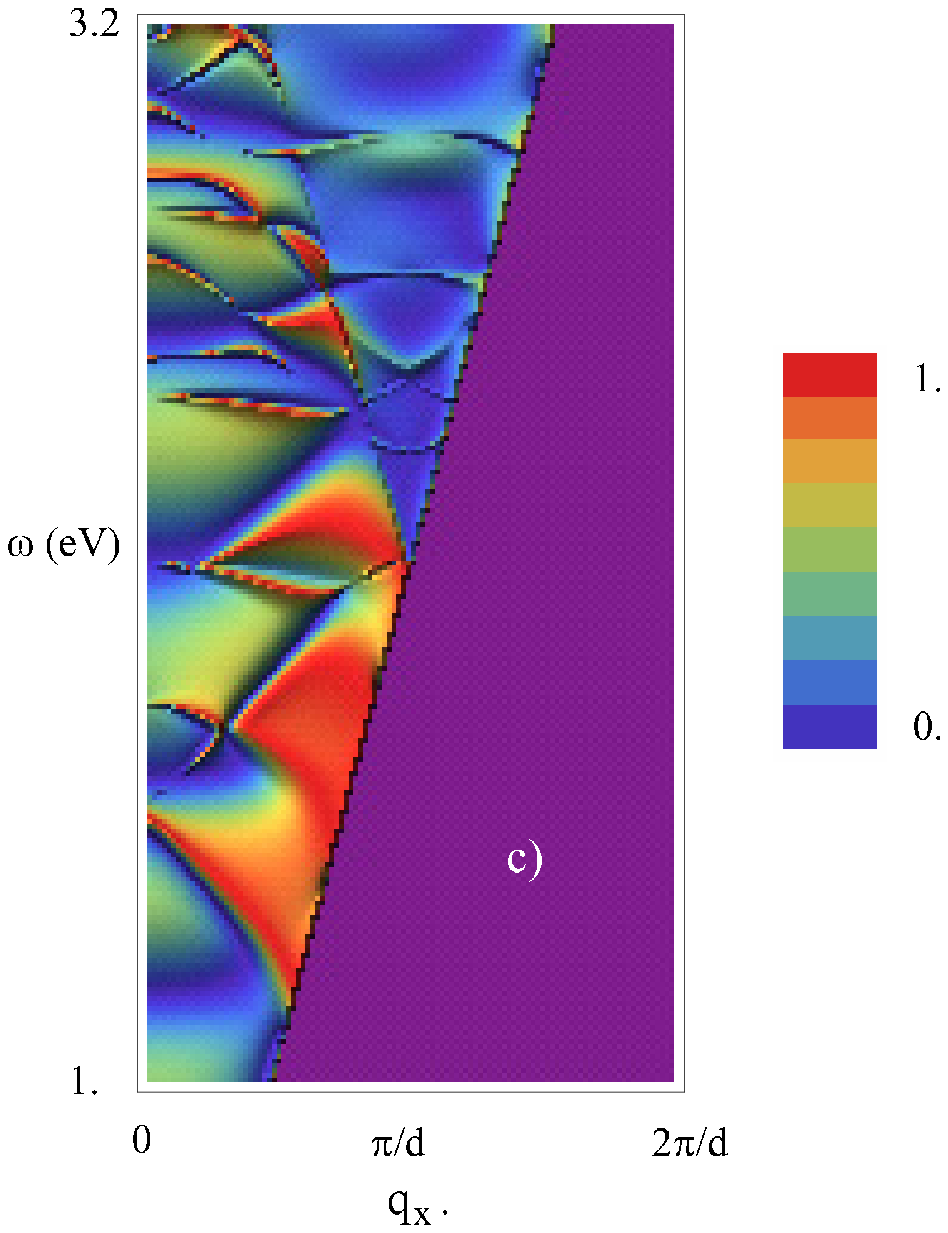}}
  \caption{Reflection intensity as a function of photon energy and in plane wave vector for three different grating with filling factors values a) $f_x = 2/3, L_z=320 nm, b) f_x = 3/4, L_z=350 nm, c) f_x = 4/5, L_z=380 nm$ }
\end{figure*}

Also in this range of photon energies is easy to verify that the sum of all the 
propagating reflection and transmission waves equals 1.

It is worth noting that, in the energy range close to the first diffracted wave escape into the vacuum, the deflected ray shows the highest intensity with transmission close to the unity ($ \sim 94$\% $
$ for $\hbar \omega  = 2.3eV$), while the zero order reflected and transmitted waves are drastically suppressed. Moreover the -1st diffracted wave, with $\textbf q_{//}  = (q_x  - G)\,\hat i$ propagates in a left direction with an energy dependent output angle $\vartheta _{out} $ as shown in Fig.7 for all the energy range of the optical spectrum. For example, at photon energy $\hbar \omega  = 2.3eV$
 the deflection angle is $\varphi _{ - 1}  \approx 20^o $, therefore an incident wave will be deflected with an angle as large as $ 50^o $, as shown in the inset of Fig.7, and intensity greater than 90$\%$ as shown in Fig.6c). The transmittivity computed considering only the $G=0,\pm 2\pi /d$ components in the dielectric tensor, shown with a dot-dashed green line in Fig.6c) ,has a rather different behaviour respect to the full calculation. 

In conclusion, we would like to underline that the negative propagation is produced by the Bragg effects in a non magnetic material with positive bulk dielectric function\cite{Luo}, a completely different effect with respect to that hypothesized by Veselago many years ago\cite{Veselago}, due to the negative dielectric and magnetic susceptibilities of the sample. Moreover, because this effect is not present in the engraved structure under the same scattering conditions, we can conclude that also for this property the higher order diffracted waves are crucially important.

\subsubsection{General scattering geometry}

In order to go a bit deeper in studying the giant reflection band and the negative propagation present in two different and characteristic energy zones close to the first and the second Wood's anomalies, let us consider the scalability properties of the SSGS.   

As discussed in ref.[11] the giant reflection band is scalable in energy if the filling factor $f_x  = L_x /d $ and the ratio $L_z /L_x $ are taken constant. For the present parameter values and $f_x  = 3/4$, $L_z$  scales according to the wavelength of the equivalent slab model, computed at the first photon threshold energy ($L_z  \approx \lambda $) of the grating as shown in Tab.1.

\begin{table}[ht]
\caption{Grating parameters}
\begin{tabular}{|c|c|c|c|c|c|}
\hline
$\omega _o (eV)$ &$\Delta \omega (eV)$ &$L_x(nm)$ &$d(nm)$ &$L_z(nm)$ &$\lambda(nm)$
  \\
\hline
1.818            & 1.273     & 150   & 200   & 234.6  & 234.588     \\
\hline
1.212         & 0.848     & 225   & 300    & 351.9    & 351.882        \\
\hline
0.909           & 0.636     & 300   & 400  & 469.2   & 469.176   \\
\hline
0.454         & 0.318     & 600   & 800   & 938.4  & 938.352   \\
\hline
\end{tabular}
\end{table}

While the filling factor determines the effective dielectric function value of  the grating, the ratio $L_z /L_x $ is  connected with the Mie energy states of the elementary nano-particle of the system; therefore, the two basic ratios  $L_x /d $ and $L_z /L_x $ completely determine the building up of the quasi-guided waves in  the former dielectric grating slab. Notice that the giant reflection band can be as large as the whole visible energy range, and moreover, it scales from infrared to ultraviolet energies according to the simple relationship: $\Delta \omega /\omega _o  \approx L_x /d  \approx 3/4$ and $L_z /L_x  \approx 3/2$, where the band width $\Delta \omega$ and the lower energy band edge $\omega _o $ are calculated for reflectivity values 1/2. Notice that for R$\geq$ 98$\% $ the band ratio is $\Delta\lambda/\lambda\geq$ 40$\% $ where $\lambda$ is calculated at the band center.

The former properties are not restricted to the parameter values considered in Tab.1.
In fact, in Fig. 8a)-8c) we plot the reflection intensity as a function of photon energy and in plane wave vector for three different grating with filling factors values  $L_x /d = p/(p+1)$ where p=2,3,4; the ratio 1/2 is not taken into account since it does not give a real giant reflection band, due to the small value of the effective dielectric constant that cancels the strong coupling condition among different in plane wave vectors components. The grating periodicity is taken constant ($d$=300nm), the wire width scale as: $L_x  = p\,d/(p+1)$ and the slab thickness ($L_z /L_x  \approx 3/2$) is optimized in order to obtain a reflection zone as large as that observed for the sample with  $L_x /d = 3/4$.  In fact rather similar zones of high reflectivity (in red in the plots), for the three different filling factor values, are present and in energy cover the range from the first to the second diffraction thresholds.

We remind that in all the three different cases the dielectric matrix of the grating is symmetrical and translational invariant with respect to its principal diagonal, and indeed has zero value for the up and down diagonal matrix elements far from the principal one with periodicities m, where m is the denominator of the filling factor: $L_x /d  = n/m$ (see eq.2). Notice that the simmetry strongly reduces the number of non zero matrix elements of the block dielectric tensor $(2N + 1) \times (2N + 1)$ to $2N(m - 1)/m$ reducing the interaction among different in-plane wave components. 
Since the giant reflection band in P polarization is a robust property of the dielectric grating slab under resonance conditions, it can be observed for filling factor values in all the range 0.5 $\div$ 1.0.

The folding of dispersion curves at $\hbar \omega  = $ 2.06 eV photon energy, at the boundary of the first Brillouin zone, is clearly shown in Fig.8; this energy is close to the escape energy into the vacuum of the first diffracted wave (for $\textbf q_{//}  = (q_x  - G)\,\hat i$) (see Fig.2).  Moreover, for photon energy of the fundamental wave (G=0) greater than 2.1eV a very small reflection intensity is observed for all the three dielectric grating slabs, as observed before  in the sample $p/(p+1)=3/4$ at $\vartheta  = 60^o $ incidence angle (see Fig.6b-6c).
 
In particular in Fig.9 reflection and transmission intensities are shown for the two electromagnetic waves travelling in the vacuum (namely: G=0 and $G =  - 2\pi /d $) in the energy range $2.1 \div 3.1\,eV$ in the folded second Brillouin zone.
\begin{figure}[t]
\includegraphics[scale=0.64]{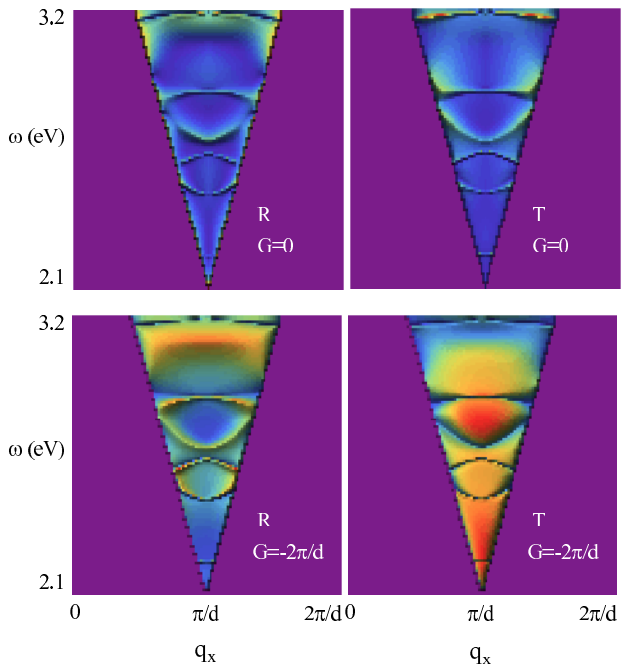}
\caption{(color online) Reflection (R) and transmission (T) intensities of the two electromagnetic waves travelling in the vacuum: G=0 and $G =  - 2\pi /d $ in the energy range $2.1 \div 3.2\,eV$ in the second Brillouin zone. The same color scale of Fig.8 is used.}
\label{fig:}
\end{figure}

We observe high intensity value only for the -1st transmission wave, therefore the optical behaviour of the dielectric grating slab is similar to a negative propagation for a rather broad energy range and incidence angles. In fact, in the second Brillouin zone, the intensity of transmitted and reflected optical waves are negligible small except for the transmission of deflected ray at $\textbf q_{//}  = (q_x  - G)\,\hat i$, and this behaviour is in agreement with the optical spectra of Fig.6c).

In Fig.10a) and 10b) the reflectivity maps for TE and TM polarization respectively are computed for a $p/(p+1)=3/4$ grating slab and angle of incidence $\vartheta  = 60^o $, as a function of $0 \leq \varphi _o  \leq \pi /2$ angle (see Fig.1); the reflection spectra for $\varphi _o  = 0$
 are also shown for sake of comparison. In the central maps are given the reflection intensities computed with only the first three waves (N=1) in the Rayliegh expansion of the electric field, while the right maps are at convergence.

\begin{figure}[b]
\includegraphics[scale=0.4]{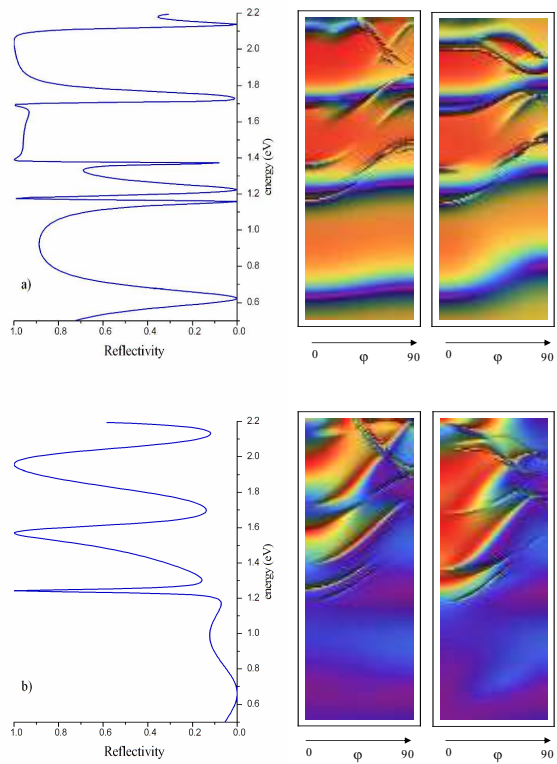}
\caption{(color online) Reflectivity maps for a) TE and b) TM polarization for a $f_x=3/4$ grating slab, angle of incidence $\vartheta  = 60^o $, as a function of the incidence plane angle $ \varphi _o $ (see Fig.1). The blu curves on the left, corresponding to the reflection spectra for $\varphi _o  = 0$, as well as the central maps are obtained for N=1; the right maps are at convergence. The same color scale of Fig.8 is used.}
\label{fig:}
\end{figure}

The TE polarization reflectivity shows a broad zone of high reflectivity values and the optical response is rather insensitive to the contribution of higher order evanescent waves, while for TM polarization where the role of the evanescent waves is crucially important the broad reflection band still remains very sensitive to the incidence angle values.

\section{Patterned mirror cavities}

The giant reflection band effect discussed in the previous sections suggest the use of sub-wavelenght gratings as very efficient reflectors in vertical microcavities since they allow the use of a single patterned layer, contrary to multilayer Bragg mirrors, and give wider wavelength ranges of high reflection.

Recently, a self-sustained rectangular dielectric grating was used at normal incidence angle in vertical cavity surface emitting lasers\cite{Lu,Huang}.

The interest for the former system with respect to the massive Bragg reflectors is twofolds: i) in energy a broad stop band is obtained ii) it should show a better response as a function of the temperature variation.

In Fig.11 the design of a vertical cavity is presented where the electromagnetic field confinement is achieved by the internal reflection of two patterned mirrors at distance L. The case of mirrors with the same periodicity of the lateral pattern and the same slab thickness $d_1=d_2$ is considered.
\begin{figure}[b]
\includegraphics[scale=0.9]{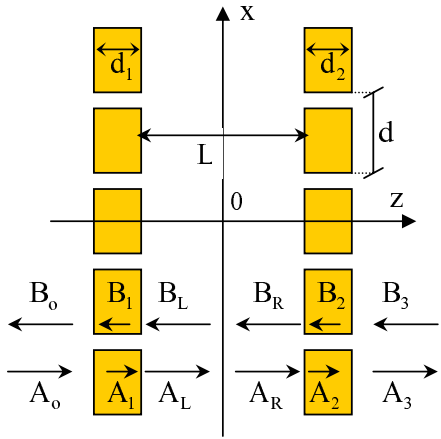}
\caption{Schematic design of a patterned cavity}
\label{fig:}
\end{figure}
If we take the in-plane wave vector along the x-axis ($\varphi_o=0$ in Fig.1) the Maxwell equations (eq.(A7) for $\varphi_o=0$) separate in S (eq.(4) and P polarization.

The in-plane electric field components in the cavity, in the grating slab and in the region to its right side, can be written as:
\begin{subequations}
\begin{align}
 0 \le z \le L /2\hspace{2.8in}&\nonumber\\
 E_{//} (r;\omega )= \sum\limits_{G } {e^{iq_x (G)x} \left[ {A_R (G)e^{iq_z (G)(z - {\textstyle{{L } \over 2}})}  + } \right.}\hspace{1.in} &\nonumber \\ 
 \left. {+ B_R (G)e^{ - iq_z (G)(z - {\textstyle{{L } \over 2}})} } \right]\hspace{1.in} \\ 
L /2 \le z \le d_2  + L /2\hspace{2.6in}\nonumber\\
 E_{//} (r;\omega ) = \sum\limits_{n} {\left[ {A_2 (n)e^{ik_n (z - {\textstyle{{L } \over 2}})} + } \right.}\hspace{1.8in}\nonumber \\ 
 \left. {+ B_2 (n)e^{ - ik_n (z - {\textstyle{{L } \over 2}})} } \right]\sum\limits_{G }^{} {\varphi _{_n } (G)e^{iq_x (G)x} } \hspace{0.8in}\\ 
d_2  + L /2 \le z \le \infty\hspace{2.8in}\nonumber\\
 E_{//} (r;\omega )= \sum\limits_{G} {e^{iq_x (G)x} \left[ {A_3 (G)e^{iq_z (G)(z - d_2  - {\textstyle{{L } \over 2}})}+ } \right.}\hspace{1.in}\nonumber \\ 
+ \left. { B_3 (G)e^{iq_z (G)(z - d_2  - {\textstyle{{L } \over 2}})} } \right]\hspace{0.8in}  
\end{align}
\end{subequations}
where the $k_n$ 2N+1 square roots of the eigenvalue problem, and the $\left\{ {\varphi _n (G)} \right\}$ correspondent 2N+1 eigenfunctions with 2N+1 G-components of the Maxwell equation (eq.A7), are computed for a chosen $q_x (0)$ value in the first Brillouin zone (-$\pi/d,\pi/d$).

A phase matrix:
\[\chi ^>
  (L) = \left( {\begin{array}{*{20}c}
   {e^{iq_z ( - N2\pi /d  )L} } & 0 & 0 & 0 & 0  \\
    \vdots  &  \ddots  &  \vdots  &  \vdots  &  \vdots   \\
   0 & 0 & {e^{iq_z (0)L} } & 0 & 0  \\
    \vdots  &  \vdots  &  \vdots  &  \ddots  &  \vdots   \\
   0 & 0 & 0 & 0 & {e^{iq_z (N2\pi /d  )L} }  \\

 \end{array} } \right)\]
match the fields in the left and right side of the cavity: 

\begin{equation}
\textbf{A}_R =\,\boldsymbol{\chi}  ^ >  (L)\textbf{A}_L \,\,\, \,\,\,\textbf{B}_R  = \boldsymbol{\chi} ^ <  (L)\,\textbf{B}_L 
\end{equation}
where the  $\chi ^ <  (L)$ matrix elements are obtained by the substitution: $e^{iq_z (G)L}  \to e^{ - iq_z (G)L} $.

By considering the S polarization, and imposing the Maxwell boundary conditions at $z = L/2$ and $z = d_2  + L/2$ we obtain the amplitudes in the transfer matrix form:

\begin{subequations}
\begin{align}
\left\{ \begin{array}{l}
 A_R (G) =  \sum\limits_n {\left[ {t_2^ >  (G,n)} \right]^{ - 1} } \varphi _n (G) \hspace{1.2in}\\
\quad \quad \quad \left\{ {A_2 (n)  + r_2^ >  (G,n)B_2 (n) } \right\} \hfill\\ 
 B_R (G) = \sum\limits_n {\left[ {t_2^ >  (G,n)} \right]^{ - 1} } \varphi _n (G)  \\
\quad  \quad \quad \left\{ {r_2^ >  (G,n) A_2 (n) +  B_2 (n)} \right\}\hfill 
 \end{array} \right.\\ 
\left\{ \begin{array}{l}
 A_3 (G) = \sum\limits_n {\left[ {t_2^ >  (G,n)} \right]^{ - 1} } \varphi _n (G) \hspace{1.2in}\\
\quad \quad \quad \left\{ {A_2 (n)e^{ik_n d_2 }  + r_2^ >  (G,n)B_2 (n)e^{ - ik_n d_2 } } \right\} \hfill\\ 
 B_3 (G) = \sum\limits_n {\left[ {t_2^ >  (G,n)} \right]^{ - 1} } \varphi _n (G)  \\
\quad  \quad \quad \left\{ {r_2^ >  (G,n)\,e^{ik_n d_2 } A_2 (n) + e^{ - ik_n d_2 } B_2 (n)} \right\}\hfill 
 \end{array} \right. 
\end{align}
\end{subequations}
where:
\begin{equation}
r_2^ >  (G,n) = \frac{{q_z (G) - k_n }}
{{q_z (G) + k_n }} \;\;\;\; t_2^ >  (G,n) = \frac{{2q_z (G)}}{{q_z (G) + k_n }}
\end{equation}
are the interface reflection and transmission amplitudes of the right grating slab.

It is however well known that, as discussed in ref. [14,19,20], the inevitable existence of evanescent solutions in dielectric gratings makes the transfer-matrix calculation numerically unstable very quickly, so that the scattering matrix formalism must be used.

In the scattering matrix form the equations 7a) and 7b) that now read:
\begin{subequations}
\begin{align}
\left\{ \begin{array}{l}
 A_R (G) =  \sum\limits_n {\left[ {t_2^ >  (G,n)} \right]^{ - 1} } \varphi _n (G) \hspace{1.2in}\\
\quad \quad \quad \left\{ {A_2 (n)  + r_2^ >  (G,n)B_2 (n) } \right\} \hfill\\ 
B_3 (G) = \sum\limits_n {\left[ {t_2^ >  (G,n)} \right]^{ - 1} } \varphi _n (G)  \\
\quad  \quad \quad \left\{ {r_2^ >  (G,n)\,e^{ik_n d_2 } A_2 (n) + e^{ - ik_n d_2 } B_2 (n)} \right\}\\ 
 \end{array} \right.\\ 
\left\{ \begin{array}{l}
 A_3 (G) = \sum\limits_n {\left[ {t_2^ >  (G,n)} \right]^{ - 1} } \varphi _n (G) \hspace{1.2in}\\
\quad \quad \quad \left\{ {A_2 (n)e^{ik_n d_2 }  + r_2^ >  (G,n)B_2 (n)e^{ - ik_n d_2 } } \right\} \hfill\\ 
 B_R (G) = \sum\limits_n {\left[ {t_2^ >  (G,n)} \right]^{ - 1} } \varphi _n (G)  \\
\quad  \quad \quad \left\{ {r_2^ >  (G,n) A_2 (n) +  B_2 (n)} \right\}\hfill 
 \end{array} \right. 
\end{align}
\end{subequations}
give the input and output field amplitudes of the right grating slab.

The forward optical response of the dielectric grating slab, for a given incident in-plane wave vector in the extended Brillouin zone $\textbf{q}_{//}   = \left[ {q_x (0) + G'} \right]\hat i = q_x (G')\,\hat i$, defined by the assumption:

F)  $A_R (G) \to \delta _{G,G'}$, $B_R (G) \to r_2 (G,G')$, $A_3 (G) \to t_2 (G,G')$ and $B_3 (G) \to 0.0$

makes the system of eq. 9a) an heterogeneous algebraic system that can be solved with respect to the values of the internal electric field amplitudes $\left\{ {A_2 (n),B_2 (n)} \right\}$.

Then the system of eq.9b) gives the matrices of forward reflection $r_2 (G,G')$ and transmission $t_2 (G,G')$ whose dimension is $(2N + 1) \times (2N + 1)$.

Due to the optical simmetry of the grating, the backward optical response, defined as:

B) $A_R (G) \to 0.0$, $B_R (G) \to t_2^{} (G)$, $A_3 (G) \to r_2^{} (G)$ and $B_3 (G) \to \delta _{G,G'} $

give the same values of reflection $r_2 (G,G')$ and transmission $t_2 (G,G')$ amplitudes of the forward one.

We can then write the scattering matrix of the system as a $2 \times 2$
 block matrix in $\left\{ {G,G'} \right\}$ in the form:

\begin{equation}
\left( {\begin{array}{*{20}c}
   {\textbf {A} _3 }  \\
   {\textbf {B} _R }  \\

 \end{array} } \right) = \textbf {S} _2 (\omega )\left( {\begin{array}{*{20}c}
   {\textbf {A} _R }  \\
   {\textbf {B} _3 }  \\

 \end{array} } \right) = \left( {\begin{array}{*{20}c}
   {\textbf {t} _2 (\omega )} & {\textbf {r} _2 (\omega )}  \\
   {\textbf {r} _2 (\omega )} & {\textbf {t} _2 (\omega )}  \\

 \end{array} } \right)\left( {\begin{array}{*{20}c}
   {\textbf {A} _R }  \\
   {\textbf {B} _3 }  \\

 \end{array} } \right)
\end{equation}

An analogous relation holds for the left grating slab:
\begin{equation}
\left( {\begin{array}{*{20}c}
   {\textbf {A} _L }  \\
   {\textbf {B} _0 }  \\

 \end{array} } \right) = \textbf {S} _1 (\omega )\left( {\begin{array}{*{20}c}
   {\textbf {A} _0 }  \\
   {\textbf {B} _L }  \\

 \end{array} } \right) = \left( {\begin{array}{*{20}c}
   {\textbf {t} _1 (\omega )} & {\textbf {r} _1 (\omega )}  \\
   {\textbf {r} _1 (\omega )} & {\textbf {t} _1 (\omega )}  \\

 \end{array} } \right)\left( {\begin{array}{*{20}c}
   {\textbf {A} _0 }  \\
   {\textbf {B} _L }  \\

 \end{array} } \right)
\end{equation}

Moreover, eq. (6) now is:
\begin{equation}
\left( {\begin{array}{*{20}c}
   {\textbf {A} _R }  \\
   {\textbf {B} _R }  \\

 \end{array} } \right) = \left( {\begin{array}{*{20}c}
   {\boldsymbol{\chi} ^ >  (L)} & {\textbf {0} }  \\
   {\textbf {0} } & {\boldsymbol{\chi} ^ <  (L)}  \\

 \end{array} } \right)\left( {\begin{array}{*{20}c}
   {\textbf {A} _L }  \\
   {\textbf {B} _L }  \\

 \end{array} } \right)
\end{equation}
where the field amplitudes are given in the $\left\{ {G,G'} \right\}$ reciprocal space, and $\boldsymbol{\chi} ^\beta  (L)$ (for $\beta  =  < {\kern 1pt} ,\, > $) are diagonal matrices.

Moreover, from eqs. (11), (12) and (13) we obtain: 
\begin{equation}
\left\{ {\begin{array}{*{20}c}
   {\textbf {A} _L  = \textbf {G} ^ >  (L)\,\left[ {\textbf {t} _1 (\omega )\textbf {A} _0  + \textbf {r} _1 (\omega )\boldsymbol{\chi}  ^ >  (L)\textbf {t} _2 (\omega )\textbf {B} _3 } \right]\;}  \\
   {\textbf {B} _L  = \textbf {G} ^ <  (L)\,\left[ {\textbf {r} _{12} (\omega )\textbf {t} _1 (\omega )\textbf {A} _0  + \boldsymbol{\chi}  ^ >  (L)\textbf {t} _2 (\omega )\textbf {B} _3 } \right]}  \\

 \end{array} } \right.
\end{equation}
where the tensors $
\textbf {G} (L)
$ are:
\begin{eqnarray}
\textbf  {G} ^ >  (L) = \left[ {\textbf {I}  - \textbf {r} _1 (\omega )\textbf  {r} _{12} (\omega )} \right]^{ - 1}\nonumber \\
\textbf  {G} ^ <  (L) = \left[ {\textbf  {I}  - \textbf  {r} _{12} (\omega )\textbf  {r} _1 (\omega )} \right]^{ - 1}
\end{eqnarray}
with $\textbf {r} _{12} (\omega ) = \boldsymbol{\chi} ^ >  (L)\textbf {r} _2 (\omega )\boldsymbol{\chi}^ >  (L)$.

The total scattering matrix of the cavity is then:
\begin{equation}
\begin{gathered}
  \left( {\begin{array}{*{20}c}
   {\textbf {A} _3 }  \\
   {\textbf {B} _0 }  \\

 \end{array} } \right) = \left( {\begin{array}{*{20}c}
   {\textbf {t} _2 (\omega )\boldsymbol{\chi} ^ >  (L)} & {\textbf {0} }  \\
   {\textbf {0} } & {\textbf {t} _1 (\omega )}  \\

 \end{array} } \right)\left( {\begin{array}{*{20}c}
   {\textbf {A} _L }  \\
   {\textbf {B} _L }  \\

 \end{array} } \right) +  \hfill \\
  \quad \quad \quad \quad \quad  + \left( {\begin{array}{*{20}c}
   {\textbf {0} } & {\textbf {r} _2 (\omega )}  \\
   {\textbf {r} _1 (\omega )} & {\textbf {0} }  \\

 \end{array} } \right)\left( {\begin{array}{*{20}c}
   {\textbf {A} _0 }  \\
   {\textbf {B} _3 }  \\

 \end{array} } \right) \hfill \\ 
\end{gathered} 
\end{equation}

Notice, that the tensor $\textbf {G} (L)$
embodies the poles of the resonant matrix, and, since the poles of the matrix are the zeros of its inverse, these are the eigen-energies of the electromagnetic field confined between the two grating slabs in the cavity. Moreover, the reduced scattering matrix embodies also the optical properties of the isolated grating slabs, as shown in the second term of the right side of eq.15). 

\begin{figure}[t]
\includegraphics[scale=0.6]{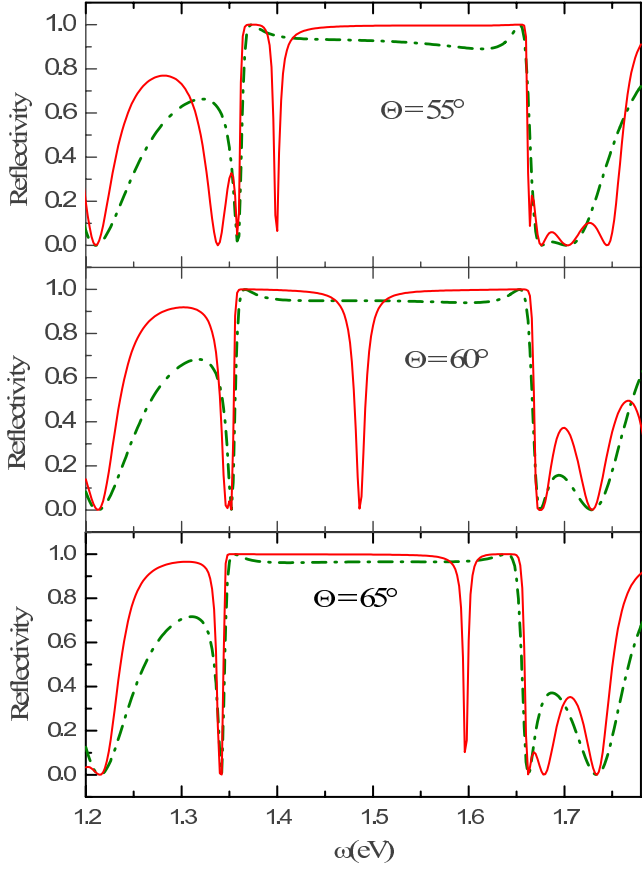}
\caption{S-polarized reflection spectra for a laterally patterned planar cavity obtained with two parallel dielectric grating slabs of filling factor $f_x  = 3/4$, elementary cell ratio  $L_z /L_x  = $
3/2 and periodicity d=300nm for incidence angle a) $\vartheta  = 55^o$, b) $\vartheta  = 60^o$, c) $\vartheta  =65^o$.}
\label{fig:}
\end{figure}

The S-polarization reflection spectra of a laterally patterned planar cavity obtained with two parallel dielectric grating slabs at distance $L $, filling factor $f_x  = L_x /d = 3/4$, elementary cell ratio  $L_z /L_x  = $3/2, and periodicity d=300nm are shown in Figs.12 a)-c) for three different incidence angles $\vartheta  = 55^o ;\,60^o ;\;65^o $.  The cavity thickness ($L = $ 826.34 nm ) is chosen for obtaining  a $\lambda /2$-cavity in a guided geometry. The reflection spectra of a single patterned dielectric grating slab are also shown in the same pictures; it is interesting to note that the reflection bands of this kind of cavity is improved with respect to the reflection band of a single dielectric grating slab. Moreover, the cavity peak  shifts in energy linearly as a function of the incidence angle, due to the small angle range chosen, and the half-width at half height is $\Gamma\approx 5meV$. 

In conclusion, these results underline that self-sustained dielectric gratings are well suited for obtaining patterned planar cavities in guided configuration; a more complete analysis will be performed in a subsequent paper.

\section{Conclusion}
The physical effects that are at basis of giant reflection band and negative light propagation in a self-sustained rectangular dielectric grating slab are clarified by comparison with an analogous optical system obtained by an homogeneous planar dielectric waveguide with a thin grating engraved on one of the surfaces\cite{rosen}. Notice, that the latter system is usually suggested in the literature\cite{rosen} as a prototype for explaining the SKWA in a dielectric grating. The role of the strong coupling among electromagnetic waves with different in plane wave vectors on the SKWAs and of the symmetry of the dielectric grating tensor are also discussed. The different physical effects that are at basis of broad reflection bands in S and P polarization are clarified. We pointed out that in a self-sustained grating slab, under resonant conditions, giant reflection bands and negative transmission can be obtained in P polarization, due to the interplay among travelling, guided and evanescent/divergent waves not very far from the Brewster angle. The optical behaviour for a general scattering geometry and mixed polarization is also briefly discussed.
Finally, the attitude on the confinement of the electromagnetic field on laterally patterned planar cavities is also discussed for S polarization.

\appendix
\section{Optical response of a self-sustained dielectric grating}
A schematic illustration of a self sustained dielectric grating slab with general incidence plane wave of $\hbar \omega $ energy is given in Fig.1.

The electric field components ($\alpha  = x,y,z$) in the grating region, $0 \leq z \leq L_z $, expanded in plane waves, are:
\begin{equation}
E_\alpha  (\boldsymbol{\rho} ,z;\omega ) = \sum\limits_{G} {E_\alpha  (\textbf{q}_{//}  + G \hat i,k;\omega )\,e^{i(\textbf{q}_{//}  + G \hat i) \cdot \boldsymbol{\rho} } } \,e^{ikz} 
\end{equation}
where $\boldsymbol{\rho} $ is the in-plane vector ($\boldsymbol{\rho} = \left( {x,y,0} \right)$).

The wave vector of the incident electric field $\textbf{q}= (\textbf{q}_{//}(0) ,q_z (0))$
has $\textbf{q}_{//}  = q_x \hat i + q_y \hat j = q_o \left[ {\cos \varphi _o\hat i + \sin \varphi _o \hat j} \right]$ and z-component $q_z (0) = \sqrt {\frac{{\omega ^2 }}
{{c^2 }}\varepsilon _o  - q_o^2 } $ with $q_o  = \frac{\omega }
{c}\sqrt {\varepsilon _o } \sin \vartheta _o$, while the in plane wave vectors of reflected, transmitted and deflected fields are $\textbf{q}_{//} (G) = (q_x  + G)\hat i + q_y \hat j = q_x (G)\hat i + q_y \hat j$ and $q_z (G) = \sqrt {\frac{{\omega ^2 }}
{{c^2 }}\varepsilon _o  - q_x^2 \,(G) - q_y^2 } $.
Therefore, the polarization plane $\alpha _\ell  $
of the deflected $\ell-th$ wave component performs a solid angle $\varphi _\ell   = arctg\left( {\frac{{q_{//} \sin \varphi _o }}
{{q_{//} \cos \varphi _o  + \ell 2\pi /d}}} \right)$ with the (x,z)-plane.

Moreover, the unit vectors of the incident electric field for TE and TM  polarizations, defined by the transversality conditions, are:
\begin{subequations}
\begin{align}
\hat \varepsilon _{TE} =(- \sin \varphi _o , \cos \varphi _o,0) \hspace{1.4in}\\
\hat \varepsilon _{TM} = (\cos \varphi _o \cos \vartheta _o, \sin \varphi _o \cos \vartheta _o , \sin \vartheta _o )\,\,\,.\hspace{0.4in}
\end{align}
\end{subequations}

The Fourier transformed Maxwell equations in mixed coordinates, are:
\begin{subequations}
\begin{align}
\left[ {\frac{{\partial ^2 }}
{{\partial z^2 }} - q_y^2 } \right]E_x \left( {G,z} \right) + q_x \left( G \right)q_y E_y \left( {G,z} \right)+
\hspace{0.5in}\nonumber\\
 - iq_x \left( G \right)\frac{{\partial E_z \left( {G,z} \right)}}
{{\partial z}} =  - \frac{{\omega ^2 }}
{{c^2 }}D_x \left( {G,z} \right)\\
\left[ {\frac{{\partial ^2 }}
{{\partial z^2 }} - q_x^2 } \right]E_y \left( {G,z} \right) + q_x \left( G \right)q_y E_x \left( {G,z} \right)+
\hspace{0.4in}\nonumber\\
 - iq_y \frac{{\partial E_z \left( {G,z} \right)}}
{{\partial z}} =  - \frac{{\omega ^2 }}
{{c^2 }}D_y \left( {G,z} \right)\\
iq_x \left( G \right)\frac{{\partial E_x \left( {G,z} \right)}}
{{\partial z}} + iq_y \frac{{\partial E_y \left( {G,z} \right)}}
{{\partial z}}+
\hspace{0.8in}\nonumber\\
 + \left[ {q_x^2 \left( G \right) + q_y^2 } \right]E_z \left( {G,z} \right) = \frac{{\omega ^2 }}
{{c^2 }}D_z \left( {G,z} \right)
\end{align}
\end{subequations}
and, given the dielectric displacement field in the form:
\begin{equation}
D_\alpha  (G,z) = \sum\limits_{G'} {\varepsilon _{G,G'}E_\alpha  (G',z)}  = e^{ikz} \sum\limits_{G'} {\varepsilon _{G,G'} E_\alpha  (G',k)} 
\end{equation}
they reduce to:
\begin{subequations}
\begin{align}
\sum\limits_{G'} {\left[ {\frac{{\omega ^2 }}
{{c^2 }}\varepsilon _{G,G'}  - q_y^2 \delta _{G,G'} } \right]} E_x \left( {G',k} \right) + \hspace{2.1in} \nonumber\\
 +q_x \left( G \right)q_y E_y \left( {G,k} \right)= k^2 E_x \left( {G,K} \right) - kq_x \left( G \right)E_z \left( {G,k} \right)\hspace{0.9in}\\
\sum\limits_{G'} {\left[ {\frac{{\omega ^2 }}
{{c^2 }}\varepsilon _{G,G'}  - q_x^2 \left( G \right)\delta _{G,G'} } \right]} E_y \left( {G',k} \right) + \hspace{1.9in} \nonumber\\
  +q_x \left( G \right)q_y E_x \left( {G,k} \right)= k^2 E_y \left( {G,k} \right) - kq_y E_z \left( {G,k} \right)\hspace{0.9in}\\
\sum\limits_{G'} {\left[ {\frac{{\omega ^2 }}
{{c^2 }}\varepsilon _{G,G'}  - \left( {q_x^2 \left( G \right) + q_y^2 } \right)\delta _{G,G'} } \right]} E_z \left( {G',k} \right) =\hspace{1.4in}\nonumber\\
 =  - k\left[ {q_x \left( G \right)E_x \left( {G,k} \right) + q_y E_y \left( {G,k} \right)} \right]\hspace{1.in}
\end{align}
\end{subequations}

The former system can be solved with respect to the z-component of the electromagnetic field:
\begin{eqnarray}
E_z \left( {G,k} \right)=\hspace{2.4in} \nonumber\\
=- k\sum\limits_{G'} {\textbf{M} _{G,G'}^{ - 1} } \left\{ {q_x \left( {G'} \right)E_x \left( {G',k} \right)+ q_y E_y \left( {G',k} \right)} \right\}
\end{eqnarray}
where the matrix of elements $
M_{G,G'}  = \frac{{\omega ^2 }}
{{c^2 }}\,\varepsilon _{G,G'}  - \left( {q_x^2 \left( {G'} \right) + q_y^2 } \right)\delta _{G,G'} $
is real and symmetric in the photonic crystal limit ($\varepsilon ''(\omega ) \to 0$).
The substitution of the z-component of the electromagnetic field in the first two equations of the system of eq.(A5) gives the Maxwell equations as a generalized eigenvalue problem:
\begin{equation}
\left( {\begin{array}{*{20}c}
   {\textbf{A} _{xx} } & {\textbf{A} _{xy} }  \\
   {\textbf{A} _{xy}^T } & {\textbf{A} _{yy} }  \\

 \end{array} } \right)\left( {\begin{array}{*{20}c}
   {\boldsymbol{\varphi}  _x }  \\
   {\boldsymbol{\varphi} _y }  \\

 \end{array} } \right) = k^2 \left( {\begin{array}{*{20}c}
   {\textbf{B} _{xx} } & {\textbf{B} _{xy} }  \\
   {\textbf{B} _{xy}^T } & {\textbf{B} _{yy} }  \\

 \end{array} } \right)\left( {\begin{array}{*{20}c}
   {\boldsymbol{\varphi} _x }  \\
   {\boldsymbol{\varphi} _y }  \\
 \end{array} } \right)
\end{equation}
where each block of the matrix, of dimension (2N+1)x(2N+1), real and symmetric in the photonic crystal approximation, is given by:
\begin{equation}
\begin{array}{*{20}c}
   {A_{xx}(G,G') = \frac{{\omega ^2 }}
{{c^2 }}\varepsilon _{G,G'} - q_y^2 \delta _{G,G'} } \hfill \\
{A_{xy}(G,G') = q_x (G')q_y \delta _{G,G'} } \hfill \\
{A_{yy}(G,G') = \frac{{\omega ^2 }}
{{c^2 }}\varepsilon _{G,G'} - q_x^2 (G')\delta _{G,G'} } \hfill \\
 {B_{xx}(G,G') = q_x (G) {\textbf{M} _{G,G'}^{ - 1}} q_x (G') + \delta _{G,G'} } \hfill  \\
 {B_{xy}(G,G')  = q_x (G) {\textbf{M} _{G,G'}^{ - 1}} q_y } \hfill  \\
    {B_{yy}(G,G') = q_y  {\textbf{M} _{G,G'}^{- 1} } q_y  + \delta _{G,G'} }\,\,\,\,\,\,. \hfill  \\
 \end{array} 
\end{equation}
We solve the generalized eigenvalue problem by the method of refs.[9,10] to which it reduces when $q_y=0$.
We first have to diagonalize the matrix
\begin{equation}
 {\textbf{A}}=\left( {\begin{array}{*{20}c}
   {{\textbf{A}} _{xx} } & {{\textbf{A}} _{xy} }  \\
   {{{\textbf{A}_{xy}^T}} } & {{\textbf{A}} _{yy} }
 \end{array} }\right)
\end{equation} 
by solving:
\begin{equation}
{\textbf{A}} \boldsymbol{\xi} _n= \lambda _n \boldsymbol{\xi} _n 
\end{equation}
where \[\boldsymbol{\xi} _n  \equiv \left( {\begin{array}{*{20}c}
   {\boldsymbol{\xi} _x }  \\
   {\boldsymbol{\xi} _y }  \\

 \end{array} } \right)_n \]
Then, trough the unitary transformation ${\textbf{U}}$ that diagonalizes ${\textbf{A}}$
we define the new quantities $\tilde{\boldsymbol{\xi _n }} \equiv {\textbf{U}} ^T \boldsymbol{\xi} _n$ and $ {\tilde {\textbf{B}} }  \equiv  {\textbf{U}} ^T {\textbf{B}}  {\textbf{U}}$
and transform the generalized eigenvalue problem into a canonical one.

In fact with the unitary transformation we obtain:
\[
\boldsymbol{\lambda} \,\tilde{\boldsymbol{\xi _n }}= k_n^2  {\tilde {\textbf{B}} } \,\tilde{\boldsymbol{\xi _n }}\]
where, $\boldsymbol{\lambda} $ is the diagonal matrix of the complex eigenvalues.

Finally, defining \[\tilde{\tilde{\boldsymbol{\xi _n }}} = \boldsymbol{\lambda} ^{1/2} \tilde{\boldsymbol{\xi _n }}\]
and \[ {\tilde {\tilde {\textbf{B}}} }  =  \boldsymbol{\lambda} ^{ - 1/2}  {\tilde { \textbf{B}} }  \, \boldsymbol{\lambda} ^{ - 1/2}\]
we obtain the canonical eigenvalue equation:
\begin{equation}
\frac{1}
{{ k_n^2 }}\,\tilde{\tilde{\boldsymbol{\xi _n }}} =  {\tilde {\tilde {\textbf{B}}} } \,\tilde{\tilde{\boldsymbol{\xi _n }}}\,\,\,.
\end{equation}

Notice that matrix $\textbf{A}$ and the renormalized matrix $\tilde {\tilde {\textbf{B}}} $ are symmetrical and real in the photonic crystal limit, while they are not Hermitian for a complex dielectric bulk value.
For $\varepsilon ''(\omega ) \ne 0$ the eigenvalues become complex and the eigenvectors are no more orthogonal. In this case, the eigenvalue problem, eq. (A7), must be solved togheter with its Hermitian conjugate and the bi-orthogonality relations between eigenvectors corresponding to complex conjugate eigenvalues must be used.

Finally, imposing the continuity of the in-plane electric and magnetic fields at z=0 and $z = L_z $ interfaces we compute the optical response of the dielectric grating slab\cite{fenn}.

In the particular case of $\varphi_o= 0$ the off-diagonal blocks of both $\textbf{A}$ and $\textbf{B}$
are zero and $\textbf{B}_{yy}$ reduces to the identity matrix. Consequently the system in eq.(A7) separates in two eigenvalue problems for S and P polarization.


\begin{references}

\bibitem{wood}
R. W. Wood, Philos.\ Mag. \textbf{4}, 396, (1902)
 
\bibitem{Rayleigh}
J. W. S. Rayleigh, Philos.\ Mag. \textbf{14}, 60 (1907) 

\bibitem{Dicke}
R. H. Dicke, Phys.\ Rev. \textbf{93}, 99 (1954) 

\bibitem{Agranovich}
Y. M. Agranovich, O. A. Dubovskii, JETP\ Lett.\textbf{3}, 345 (1966) 

\bibitem{Yablonovich}
E.Yablonovich, Phys.\ Rev.\ Lett.\textbf{58}, 2059 (1987)

\bibitem{Luo}
Chiyan Luo, Steven G.Johnson, J.D.Joannopoulos, J.B.Pendry, Phys.\ Rev.\ B\textbf{65}, 201104 (2002)

\bibitem{John}
S. John, Phys.\ Rev.\ Lett.\textbf{58}, 2486 (1987)

\bibitem{Magnusson}
R. Magnusson, S.S.Wang, Appl.\ Phys.\ Lett.\textbf{61}, 2486 (1987)

\bibitem{rosen}
D. Rosenblatt, A. Sharon, and A. A. Friesem, "Resonant grating waveguide
structures," IEEE J. Quantum Electron. 33, 2038 (1997).

\bibitem{laura1}
L.Pilozzi, A.D'Andrea, R.Del Sole, Phys.\ Rev.\ B \textbf{54}, 10751 (1996)

\bibitem{fenn}
L. Pilozzi, A. D'Andrea and H. Fenniche, Phys.\ Rev.\ B \textbf{64}, 235319, (2001)
 
\bibitem{Lu}
C.Lu, M.C.Y.Huang,C.F.R.Mateus,C.J.Chang Hasnain, Y.Suzuki, Appl.\ Phys.\ Lett.\  \textbf{88}, 031102 (2006)

\bibitem{Huang}
Michael C.Y.Huang, Y. Zhou, Connie J. Chang-Hasnain, Nature Photonics  \textbf{1}, 119 (2007)

\bibitem{Tikhodeev}
S. G. Tikhodeev, A. L. Yablonskii, E. A. Muljarov, N. A. Gippius, and Teruya Ishihara Phys.\ Rev.\ B \textbf{66}, 045102 (2002)

\bibitem{Felbacq}
Diedier Felbacq, Maria Cristina Larciprete, Concita Sibilia,  
      Mario Bertolotti, Michael Scalora, Phys.\ Rev.\ E \textbf{72}, 066610 (2005)


\bibitem{du}
Junjie Du, Zhifang Lin, S.T.Chui, Wanli Lu, Hao Li, Aimin Wu, Zhen Sheng, Jian Zi, Xi Wang, Shichang Zou, Fuwan Gan, Phys.\ Rev.\ Lett. \textbf{106}, 203903 (2011)

\bibitem{Podolskiy}
Viktor A.Podolskiy and Evgenii E.Narimov, Phys.\ Rev.\ B \textbf{71}, 201101 (2005)


\bibitem{Veselago} 
 V.G.Veselago, Sov.\ Phys.\ USPEKHI \textbf{10}, 509 (1968)


\bibitem{Ko}
D.Y.K.Ko and  J.C.Inkson Phys.\ Rev.\ B \textbf{38}, 9945 (1988)


\bibitem{Whittaker}
D.M.Whittaker, I.S.Culshaw, Phys.\ Rev.\ B \textbf{60}, 2610 (1999)

\end{references}
\end{document}